# Time series and machine learning to forecast the water quality from satellite data

Maryam R. Al Shehhi[1]* and Abdullah Kaya[2]

*1 Civil, Infrastructure and Environmental Engineering, Khalifa University, Abu Dhabi, UAE*

*2 Industrial and Systems Engineering, Khalifa University, Abu Dhabi, UAE*

*Corresponding author*

*Maryam R. Al Shehhi (maryamr.alshehhi@ku.ac.ae)*

*Keywords: water quality, MODIS, ocean color, neural network, SARIMA, satellite, time series, forecast*

**Abstract**

Managing and protecting the quality of water for present and future generations of coastal regions should be a central concern of both citizens and public officials. Remote sensing can contribute to the management and monitoring of coastal water and pollutants. Algal blooms are a coastal pollutant that is a cause of concern. Many satellite data, such as the Moderate Resolution Imaging Spectroradiometer (MODIS), have been used to generate water-quality products to detect the algal blooms such as chlorophyll a (Chl-*a*), a photosynthesis index called fluorescence line height (FLH), and sea surface temperature (SST). It is important to characterize the spatial and temporal variations of these water quality products by using the mathematical models of these products. However, for monitoring, enforcement, and legislative purposes, environmental protection agencies or pollution control boards will need nowcasts and forecasts of any pollution. Therefore, we aim to predict the future values of the MODIS Chl-*a*, FLH, and SST of the water. This will not be limited to one type of water but, rather, will cover different types of water varying in depth and turbidity. This is very significant because the temporal trend of Chl-*a*, FLH, and SST is dependent on the geospatial and water properties. For this purpose, we will decompose the time series of each pixel into several components: trend, intra-annual variations, seasonal cycle, and stochastic stationary or irregular parts. We explore three such time series machine learning models that can characterize the non-stationary time series data and predict future values, including the Seasonal ARIMA (Auto Regressive Integrated Moving Average) (SARIMA), regression, and neural network. The results indicate that all these methods are effective at modelling Chl-*a*, FLH, and SST time series and predicting the values reasonably well. However, regression and neural network are found to be the best at predicting Chl-*a* in all types of water (turbid and shallow). Meanwhile, the SARIMA model provides the best prediction of FLH and SST.

1. Introduction

Ocean color remotely sensed data have been widely used to monitor the water quality of oceans. This monitoring has been achieved by modelling the water quality parameters based on the satellite radiative transfer models and empirical models. However, while radiative transfer models are complex, the preference is for empirical models. Empirical models can successfully estimate water quality parameters. For example, the empirical algorithm Ocean Colour 3 (OC3) developed by [1] in 1998 has been intensively used to map chlorophyll a (Chl-*a*), an indicator of algal growth in the ocean, in several offshore regions of Chesapeake Bay, West Florida, the Arabian Sea, Korea, and China [2]–[6][7]. In addition, the fluorescence line height (FLH) algorithm developed by [8] was able to estimate algal growth based on fluorescence and could differentiate between algae and sediments in the ocean [9][10][11][12][13][14][15]. Sea surface temperature (SST), which is another indicator of ocean algal productivity, has also been successfully retrieved from ocean colour satellites based on the thermal band algorithm [16].

All these aforementioned satellite-based modelled parameters can be used to determine the water quality of the ocean and detect marine disasters. Harmful algal blooms (HABs) are among the most common marine disasters that could be monitored from space. "HABs" refers to the overgrowth of algal species in water. These species can transform the water color to red, green, or brown. They can cause death and/or poison fish and birds (which affects the food cycle), affect human health by irritating the eyes and the respiratory system, damage/coastal industrial plants, and affect tourism. HABs have occurred worldwide in several regions such as West Florida, the Gulf of Mexico, China, and the Arabian Gulf [17]–[20][21].

Forecasting the presence of these algal blooms is necessary to monitor water quality and protect the aquatic system and coastal industrial plants. The time series and machine learning are excellent approaches for estimating satellite water quality parameters and forecasting their trends in the future. Several studies have explored the time series of Moderate Resolution Imaging Spectroradiometer (MODIS)-retrieved products. For instance, Xiao et al. (2011) studied the temporal variation of the leaf area index (LAI) in France, the US, and Canada [22]. They found that a Seasonal Auto-Regressive Integrated Moving Average (SARIMA) model is the best model for deriving the LAI climatology. An Ensemble Kalman Filter (EnKF) was used for the perdition which can update the biophysical variable after the new observation is obtained. Also, Reyburn et al. (2011) studied the effect of climate and ocean environmental variability (using MODIS data) on cholera incidents, especially in sub-Saharan Africa. They established that the SARIMA model can determine the effect of these factors (especially temperature and rainfall) on cholera incidents.

Despite this powerful performance of satellite images in monitoring water quality, such images have some limitations. The main limitation of remotely sensed data is the dusty weather and clouds that can mask the satellite images and limit the satellite's coverage, leading to missing data. This problem is intensified in arid and turbid regions. To explore the effectiveness of the time series techniques in forecasting water quality parameters in turbid and shallow water, we have selected the Arabian Gulf as a case study in this paper. The Arabian Gulf is one of the shallowest waterbodies (depth<100m) where HABs occur. The water of this region is very turbid. Blooms usually occur during the winter and spring in this region. The worst documented HAB event that occurred in this region took place in 2008-2009. The blooms of this event spread over this region for over nine months and killed thousands of tons of fish. Due to the high frequency of algal bloom occurrence in this turbid region, there is an urgent need to monitor the region's water quality from space and forecast the occurrence of blooms.

This paper explores three statistical methods—Seasonal Autoregressive Integrated Moving Average (SARIMA), neural network, and regression—to forecast the main water quality parameters mentioned earlier (Chl-*a*, FLH, and SST) in turbid water. These techniques have rarely been used in the ocean remote sensing field in shallow and turbid water. Thus, the main objectives of this study are to: 1) explore the applicability of these three methods to modelling MODIS Chl-*a*, FLH, and SST time series in shallow and turbid water and forecast their future values and 2) explore approaches to overcoming the problem of missing satellite data.

2. Data Analysis

The Arabian Gulf is a semi-enclosed sea open to the Sea of Oman from the east. This bay is shaped like the letter 'S', but horizontally. It is characterized by different properties from east to west. For example, its depth decreases from the east, the Sea of Oman side, from 4000 meters towards the north to less than 50 meters, while turbidity increases from east-south to west-north by over 5 meters of Secchi disk depth (SDD) [23]. SDD is an indication of water turbidity in which high values of SDD indicate clear water and low values of SDD indicate turbid water. Due to the high geo-variations of turbidity in the Arabian Gulf region, it is interesting to study the water quality parameters at variable levels of depth and turbidity and the long-term pattern in the Gulf region. For this purpose, and considering this contrast between east and west, we chose three sites for studying the temporal changes in water quality parameters (Chl-$a$, FLH, and SST) temporarily and spatially. These sites are in: 1) deep (depth>100m) and less turbid (SDD>15m) water; 2) shallow (50<depth<100m) and turbid (SDD<10m) water; and 3) very shallow (depth<50m) and turbid (SDD<10m) water, as shown in Figure 1.

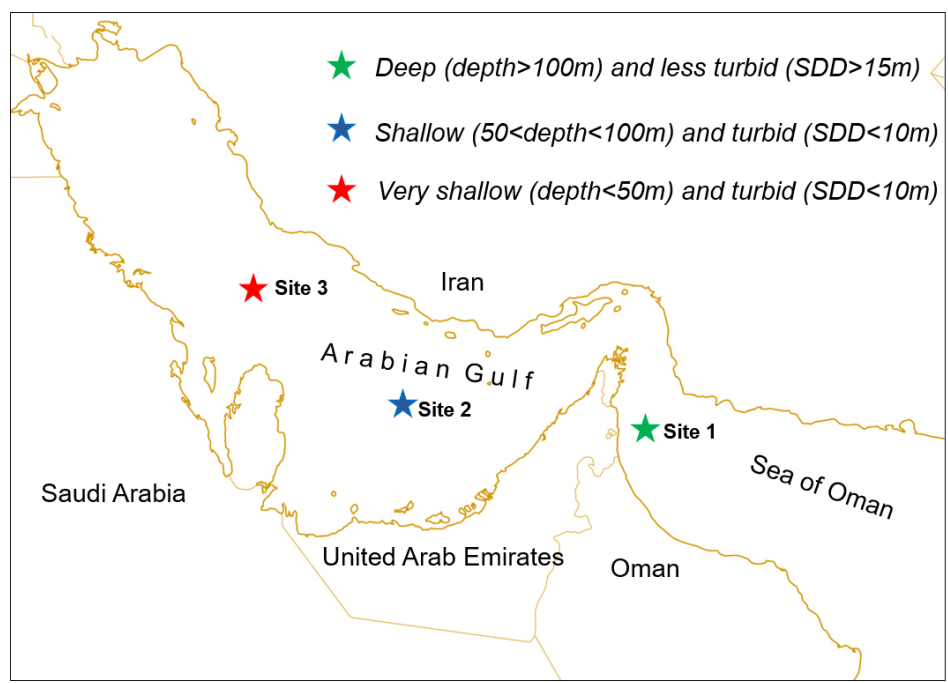

Figure 1. Map of the Arabian Gulf showing the three sites selected of varying turbidity and depth.

The data for Chl-$a$, SST, and FLH were collected over 10 years from 2003 to 2012 at these three locations. These data were obtained from MODIS, which is a NASA Earth Observing System (EOS). It is a major instrument onboard the Terra (EOS AM) and Aqua (EOS PM) satellites, launched in December 1999 and May 2002, respectively. These sensors have been extensively used to monitor fluctuating water quality in different regions such as Florida, China, Mexico, and Korea.

The data extracted must be initially analysed for stability over time and the amount of missing data before time series modelling is performed to ensure the quality of the data used in the modelling. For this purpose, we analysed the data pattern, heteroskedasticity, and missing data as follows.

- *Data patterns and stability*

Figure 2 shows the monthly data of Chl-*a*, SST, and FLH retrieved from MODIS between 2003 and 2012 at the three sites described above. As shown in this figure, Chl-*a* data fluctuated seasonally between 2003 and 2012 in the three different locations. At Site 3 (shallow-turbid), the concentration of Chl-*a* shows decreased behavior throughout the entire period. However, in an annual comparison, the concentration of Chl-*a* is higher in the winter (January, February, and March) as compared to the summer. The highest concentration of Chl-*a*—around 12 mgm$^{-3}$—was observed in 2003, while the peak values gradually decreased as shown in Figure 2. Likewise, at Site 2 (deep-turbid), the concentration of Chl-*a* shows fluctuating behavior. The highest level of Chl-*a* was around 4 mg$^{-3}$, which was also observed in 2003 (Figure 2). Another peak of Chl-*a* was observed in 2007. The same behavior at Site 2 was observed at Site 1 (deep and less turbid) with the highest concentration of Chl-*a* observed in 2007 and 2008 (Figure 2). Likewise, FLH data fluctuated over the study period (Figure 2). No high peaks are observed, but seasonality is noted, especially at Site 1. However, SST shows very clear seasonal behavior. It increases in the summer (35°C) and decreases in winter (19°C) as shown in Figure 3. Table 1 shows the statistics of the variables of Chl-*a* (mgm$^{-3}$), SST (°C), and FLH.

The patterns of these data shown in Figure 2 are not adequate for investigating the heteroskedasticity of the data. Heteroskedasticity occurs when the residuals are not identical at all observations, which is frequently encountered in regression analysis. If heteroscedasticity exists, logarithmic transformation is one of the techniques that can be used to overcome heteroskedasticity. The advantages of this technique are: 1) it shifts the contrast structure from heterogeneous to homogenous and 2) it does not affect the complexity of the model.

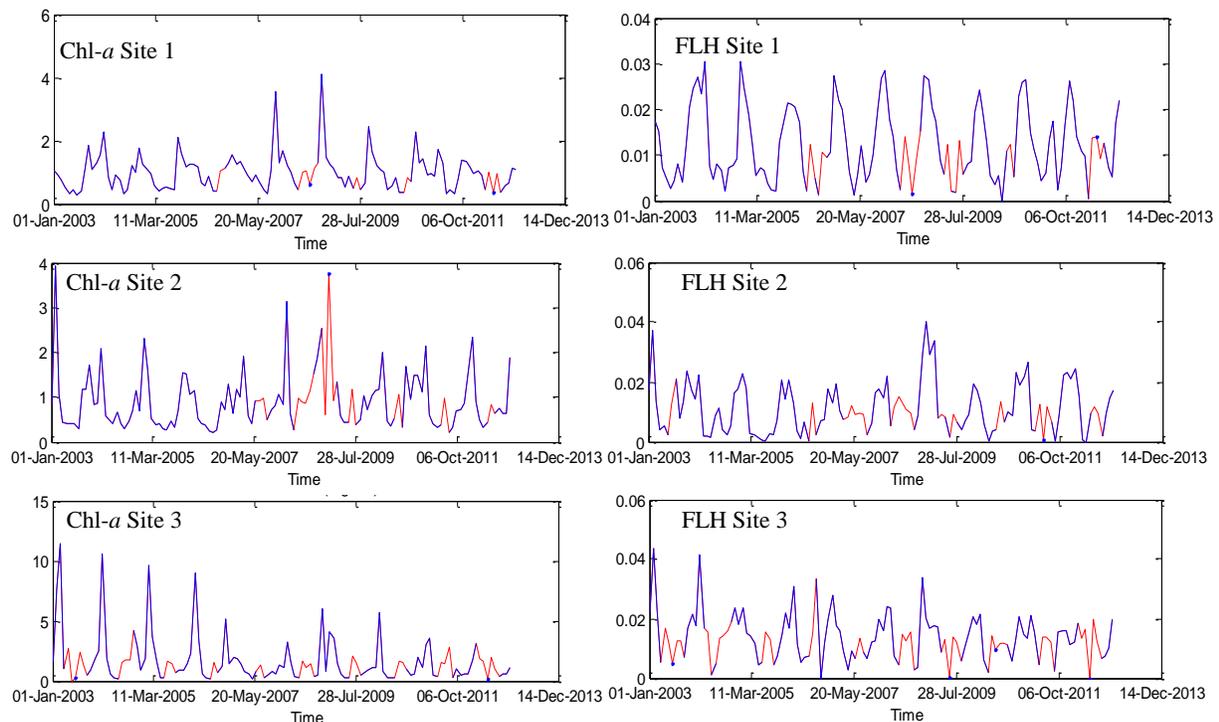

Figure 2. Time series of Chl-*a* and normalized Fluroscence line height (FLH) at three different locations (red colour refers to the missing data).

Table 1. The summary of the explanatory variables of Chl-*a* (mgm$^{-3}$), SST ($^{o}$C) and FLH.

| Parameter | Mean | Std. | Max | Min |
|---|---|---|---|---|
| Chl-*a* (3) | 1.717 | 2.138 | 11.464 | 0 |
| Chl-*a* (2) | 1.049 | 1.163 | 9.915 | 0.208 |
| Chl-*a* (1) | 0.994 | 0.635 | 4.104 | 0.294 |
| SST (3) | 27.6 | 2.670 | 31.954 | 22.662 |
| SST (2) | 27.539 | 4.540 | 34.183 | 19.864 |
| SST (1) | 25.64 | 5.396 | 33.655 | 16.480 |
| FLH (3) | 0.014 | 0.01 | 0.043 | 0 |
| FLH (2) | 0.011 | 0.009 | 0.040 | 0 |
| FLH (1) | 0.012 | 0.009 | 0.031 | 0 |

Two heteroskedasticity tests, a white test and a Breusch-Pagan test, were applied on the nine data sets to ensure that the variance did not change over time. Table 2 shows the probabilities of the white and Breusch-

Pagan test of the nine datasets. If the probability of the heteroskedasticity is lower than 0.05, there is significant heteroskedasticity.

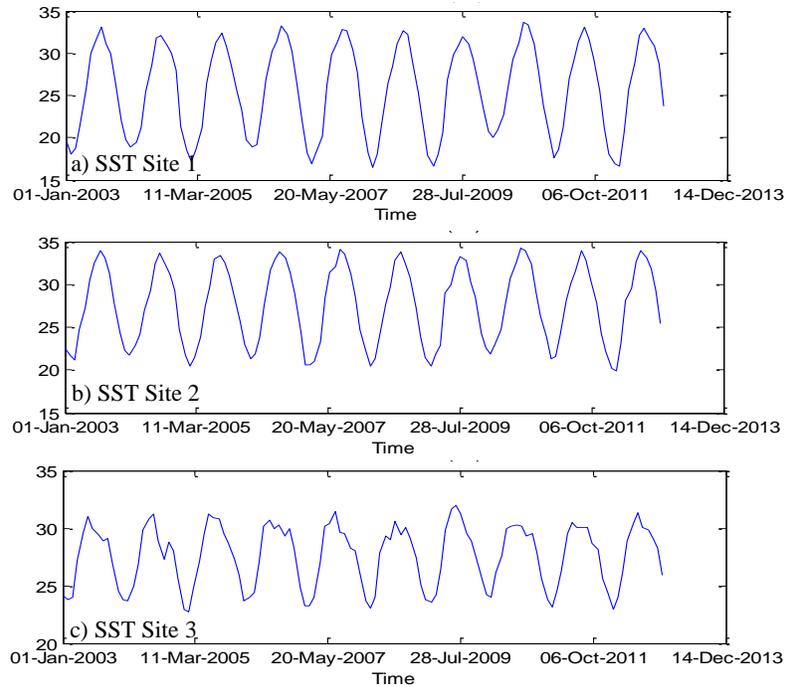

Figure 3. Time series of Sea Surface Temperature (SST) ºC at three different locations (red colour refers to the missing data).

As shown in Table 2, no heteroskedasticity was detected in all data groups except for Chl-*a* and FLH at Site 1 (very shallow-turbid), Therefore, the residuals of these two cases were plotted to determine whether the residuals were identically distributed. Figure 4 shows that the residuals are randomly distributed, without any clear trends. However, there is a slight trend in the concentration of Chl-*a*, due to the peaks of Chl-*a* in winter. Therefore, it can be claimed that those two data sets have negligible heteroskedasticity and that, as a result, there is no need to transform the nine data sets. One more important note to mention is that the high peaks of Chl-*a* observed in some months cannot be considered outliers because they are part of the same data. These high levels of Chl-*a* have been caused by the HAB events that occurred at these particular times. Removing these points would result in the loss of very important information. Therefore, it is important to maintain and make use of this data for time series modelling.

Table 2. The probabilities (p-value) of the nine data sets.

| Parameter | White test | Breusch Pagan test |
|---|---|---|
| Chl-*a* (3) | 0.0034 | 0.0000 |
| Chl-*a* (2) | 0.6282 | 0.1166 |
| Chl-*a* (1) | 0.3955 | 0.8772 |
| SST (3) | 0.8562 | 0.8022 |
| SST (2) | 0.9689 | 0.994 |
| SST (1) | 0.9428 | 0.8789 |
| FLH (3) | 0.0422 | 0.0012 |
| FLH (2) | 0.8655 | 0.6679 |
| FLH (1) | 0.3905 | 0.3138 |

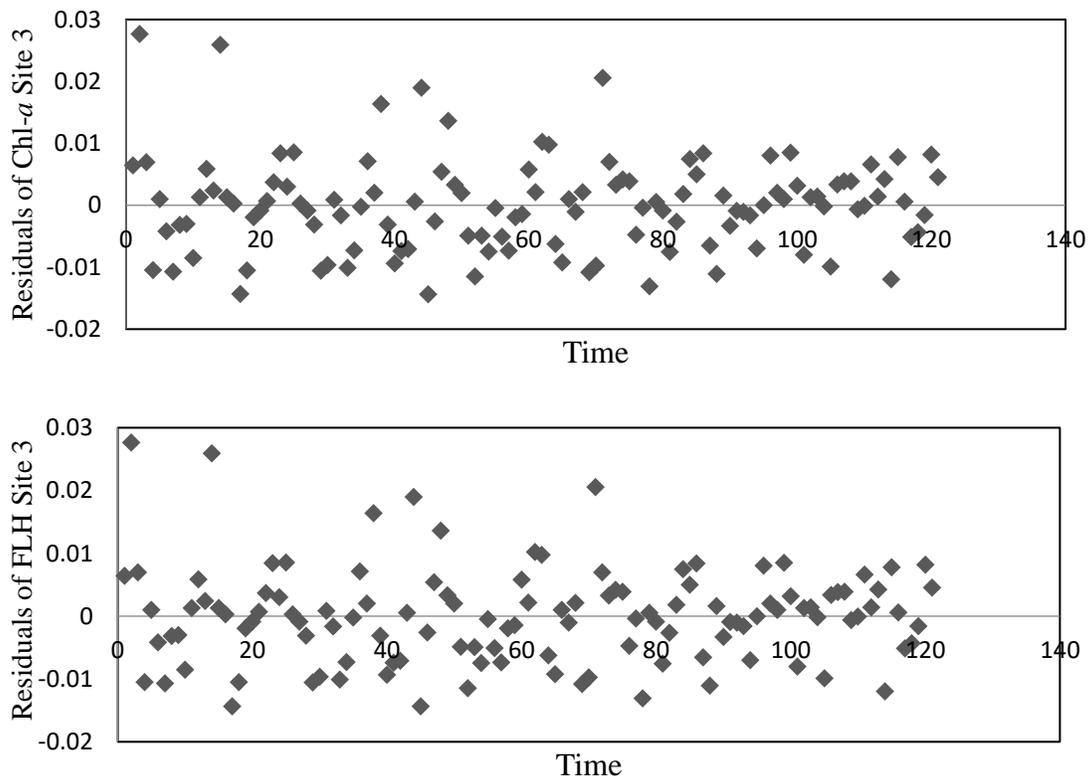

Figure 4. Residuals of Chl-*a* concentration (mg/m$^3$) and FLH at Site 3.

- *Missing data*

The red color in Figure 2 indicates the number of missing data. As shown in this figure, there is a lack of Chl-*a* and FLH data in all three sites, caused by weather conditions such as clouds and dust storms that prevented the satellite from collecting data at these particular locations and times. There are three types of missing data: 1) Missing completely at random (MCAR), 2) Missing at random, and 3) Missing not at random (MNAR). Randomly missing means that the possibility of losing Chl-*a* and FLH data is not related to the ranges of these values. In other words, external factors—such as the presence of clouds or dust—cause this data to go missing. The type of missing data is examined by the number of missing data varying greatly over the months, although the values of Chl-*a* and FLH data are still comparable. Most of the missing data are observed in the summer period, though the cause of the missing data is not due to the values of Chl-*a* or FLH data, as shown in Figure 5. For example, at Site 3, the number of missing data in June and July shows high variability, while the average Chl-*a* concentration is very similar. To further examine the relation between the number of missing data and their data ranges, the correlation between the mean values of the data (Chl-*a* and FLH) and the number of data lost has been obtained, as shown in Table 3. This table indicates that the correlation coefficient ($R^2$) is low, with values lower than 0.4. Therefore, the missing data are not MNAR and can be considered MAR or MCAR based on regression analysis. A good feature of MAR or MCAR is that missing data can be estimated without modelling the probability of item loss [24].

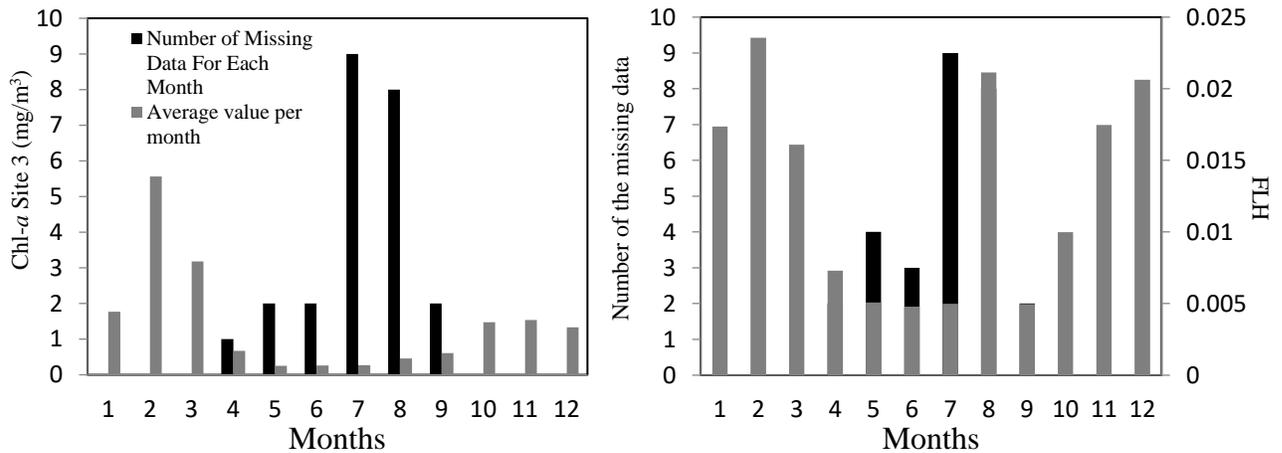

Figure 5. The number of the missing data and the average values of Chl-*a* and FLH per month.

Table 3. The results of the regrssion between the number of the missing data and the average value of the data.

| Parameter | Chl-*a* | | | FLH | | |
|---|---|---|---|---|---|---|
| Location | Site 3 | Site 2 | Site 1 | Site 3 | Site 2 | Site 1 |
| R Square | 0.227 | 0.236 | 0.393 | 0.109 | 0.246 | 0.403 |
| Lower 95.0% | -2.230 | -2.854 | -2.649 | -382.138 | -315.718 | -172.328 |
| Upper 95.0% | 0.292 | 0.335 | -0.177 | 92.707 | 0.462 | -30.681 |

3. Methods

To develop robust time series and machine learning models that can predict the parameters of Arabian Gulf water quality, it is necessary to verify the quality of the time series data by handling the problem of missing data mentioned above and then performing the time series modelling. The time series data cover three water quality parameters (Chl-*a*, FLH, and SST) retrieved monthly between 2003 and 2012. For each parameter, there are a total of 120 points. Eighty-four data points (around 70%) will be used in the modelling, while the rest (30%) will be used for validation of the model.

3.1 Handling of missing data

The missing data described in section 2 are handled by the "Multiple imputations" method, which imputes the data several times [25]. The advantage of multiple imputations is that it obtains unbiased estimates. This method can be used when the missing data are either MAR or MCAR [26]. The most common method is regression, but this method provides negative numbers for the estimated Chl-*a* values while the observed values are always positive. An alternative to this method is the mean predictive matching method (PMM), which is very similar to the linear regression method [25]. Therefore, the missing data were imputed 20 times using the Stata PMM calculation method. Then an average of 20 values was calculated and all missing data were replaced by those values that qualified for time series modelling.

3.2 Time series techniques

There are two time series and machine learning modelling approaches: univariate and multivariate. Univariate time series refers to a time series containing single observations recorded in equal time increments. If the data are equal in spacing, it is not necessary to explicitly provide the time or index variable [27]. An example of the univariate approach is the Box-Jenkins approach that will be applied in this study. In addition, the multivariate approaches of regression and neural network will be examined in this paper. The three approaches are described as follows:

- Box-Jenkins approach

The Box-Jenkins approach used in this study is the Seasonal Auto-Regressive Integrated Moving Average (SARIMA) model. The parameters of this model are shown in Eq. 1

$$SARIMA(p, d, q)(P, D, Q)_m \qquad (1)$$

Where the trend elements are p: trend autoregression order, d: trend difference order, q: trend moving average order as part of the trend elements, and the seasonal elements P: seasonal autoregressive order, D: seasonal difference order, Q: seasonal moving average order and m: the number of time steps for a single seasonal period.

The time series data should be weakly stationary to be modelled by the Box-Jenkins approach. Stationary means that the mean, variance, and autocorrelation structure of the data do not change over time. Stationary means no trend, a constant variance over time, a constant autocorrelation structure over time, and no periodic fluctuations. It is found that double differencing can improve the stationarity of the data, where the first differencing is to remove the slight trend and the second differencing is to remove the seasonality. However, differencing does not completely remove the variability and trends based on the autocorrelation function (ACF) and partial autocorrelation function (PACF) of the deseasonalized data. They show significant autocorrelation especially at lag 12, suggesting the use of AR 12 and seasonal 12 terms as a SARIMA model.

The remaining parameters of the SARIMA model are identified by the Akaike information criterion (AIC) and Bayesian information criterion (BIC). AIC and BIC deal with the trade-off between the model's complexity and goodness of fit and have been widely used for model identification in time series and linear regression. BIC has been widely used with sets of maximum likelihood-based models[28][29]. we found that both approaches, AIC and BIC, can give slightly different orders of the SARIMA model. Therefore, the outcomes of both approaches will be compared to examine the performance of each approach in identifying the best model for each parameter.

- Regression approach

Four regression models have been tested to find the best relation between the Chl-*a* concentration and the other explanatory variables: SST, FLH, and time. The regression models are expressed in Eqs. 2-5. For example, the first regression model relates the Chl-*a* concentration with time. The dummy variables are used to express each month and to overcome the problem of seasonality. The second regression model is to relate the Chl-*a* concentration with FLH (t), SST (t), and time. However, the third and fourth regression models are dependent on the variable at time (t) and (t-1).

$$\text{Chl-}a(t) = f(T(t)) \tag{2}$$

$$\text{Chl-}a(t) = f(T(t), FLH(t), SST(t)) \tag{3}$$

$$\text{Chl-}a(t) = f(T(t), FLH(t), FLH(t-1), SST(t), SST(t-1)) \tag{4}$$

$$\text{Chl-}a(t) = f(T(t), FLH(t), FLH(t-1), \text{Chl-}a(t-1)) \tag{5}$$

- Neural network approach

The neural network is an information processing technique that mimics the behavior and operation of biological nervous systems. As is known, the human brain consists of billions of neurons connected to thousands of other neurons. The neural network algorithm model is a given problem in a similar way. It consists of six main elements: the node, input links, activation function, output of a node, links to other nodes, and link weights. Many input and output links are connected to each node in the network. Each link in the network has a specific weight value assigned to it. The algorithm learns by changing the weights of the links in the network [30][30][29][28][27][26][25][25][24][25][25]. The learning process of the neural network starts with the assigning of initial weights to all links in the network. Then items in the training data set are fed into the network one by one. The output is calculated by multiplying the input value on a given link by the weight of that link. This is done to all input links. At the end, the results are summed and compared with the threshold value. If the output of the neural network does not match the actual output, the neural network modifies the link weights by either increasing or decreasing the weight value by a small amount. However, not all the weights are changed. Only the weights of the active links are affected. An active link is a link that has a non-zero input value. If the output value of the neural network matches the actual output value, the weights of the links should not be modified. It is not a good idea to stop the classifier while it is running because the error might be less than a particular threshold and it can also cause overfitting. The threshold value in the neural network is determined by the activation function. The most common activation function is the step function, which can be used to classify the data into two classes. The most common types of neural networks are single-layer and multi-layer. Single-layer networks are used to solve linear problems while multi-layer networks are used to solve non-linear and more complex problems. Backpropagation is a common method used to adjust the weights on the network links ([30]. The network consists of many nodes connected by many links. Each link in the neural network has its own weight, which results in an overall complex modelling of the problem. The neural network might not reach a stable status. In other words, different training inputs may continue affecting the weights on the links forever. Therefore, it should have a stopping point to terminate the training process [31]. The neural network can be applied for time series purposes based on two approaches: nonlinear autoregressive (NN-NAR) and nonlinear autoregressive with external input (NN-NARX). In NAR, we have considered the past values of Chl-*a*, FLH, and SST to predict the future values. Meanwhile, the NARX approach has been used only to predict Chl-*a* values based on the external factors (FLH and SST) and the past Chl-*a* values.

4. Validation of the Time Series Models

The three time series techniques described earlier have been used to model Chl-*a*, FLH, and SST at the three locations that differ in terms of turbidity and water depth. The performance of these methods is evaluated based mainly on their predictability at these locations. The predictive abilities of these three methods were tested for estimating Chl-*a*, FLH, and SST values using the prior seven years of data (2003-2010). Two measures were used to evaluate the predictions: root mean square error (RMSE) and coefficient of determination ($R^2$). To evaluate the forecasting abilities of the three methods according to water type, the performance details are examined below:

- Deep (>100m) and not-turbid water

The models—SARIMA, regression, and neural network—have been validated to estimate Chl-*a*, FLH, and SST in the deep and non-turbid water of Site 1. The coefficients and parametrization of the SARIMA and regression models found for this type of water (deep and not-turbid) are shown in Tables 4 and 5. As shown in Table 4, AIC- and BIC-based SARIMA models have mostly given comparable performances.

Table 4. Summary of the goodness of fit and RMSE of all the SARIMA models.

| SARIMA | BIC | $R^2$ | RMSE | AIC | $R^2$ | RMSE |
|---|---|---|---|---|---|---|
| Chl-*a* (3) | SARIMA(1,1,1)(1,1,1)$_{12}$ | **0.382** | 1.335 | SARIMA(4,1,1)(1,1,1)$_{12}$ | 0.359 | 1.511 |
| Chl-*a* (2) | SARIMA(3,1,1)(1,1,1)$_{12}$ | **0.535** | 0.545 | SARIMA(1,1,2) (1,1,1)$_{12}$ | 0.510 | 0.634 |
| Chl-*a* (1) | SARIMA(1,1,2)(1,1,1)$_{12}$ | 0.504 | 0.599 | SARIMA(1,1,2) (1,1,1)$_{12}$ | 0.504 | 0.599 |
| FLH (3) | SARIMA(1,1,1)(1,1,1)$_{12}$ | 0.402 | 0.007 | SARIMA(3,1,4)(1,1,1)$_{12}$ | **0.408** | 0.007 |
| FLH (2) | SARIMA(4,1,1)(1,1,1)$_{12}$ | 0.522 | 0.018 | SARIMA(4,1,4)(1,1,1)$_{12}$ | **0.718** | 0.006 |
| FLH (1) | SARIMA(3,1,4) (1,1,1)$_{12}$ | 0.621 | 0.010 | SARIMA(3,1,4)(1,1,1)$_{12}$ | 0.621 | 0.010 |
| SST (3) | SARIMA(2,1,2)(1,1,1)$_{12}$ | **0.978** | 0.564 | SARIMA(3,1,4)(1,1,1)$_{12}$ | 0.976 | 0.594 |
| SST (2) | SARIMA(1,1,1) (1,1,1)$_{12}$ | **0.986** | 1.059 | SARIMA(1,1,2) (1,1,1)$_{12}$ | 0.985 | 1.042 |
| SST (1) | SARIMA(1,1,3)(1,1,1)$_{12}$ | 0.981 | 1.364 | SARIMA(4,1,4)(1,1,1)$_{12}$ | **0.983** | 1.289 |

Table 5. Summary of the goodness of fit and RMSE of all the regression forms.

| Site | (3) | (2) | (1) | (3) | (2) | (1) | (3) | (2) | (1) | (3) | (2) | (1) |
|---|---|---|---|---|---|---|---|---|---|---|---|---|
| | Regression 1 of Chl-*a* | | | Regression 2 of Chl-*a* | | | Regression 3 of Chl-*a* | | | Regression 4 of Chl-*a* | | |
| $R^2$ | 0.435 | 0.436 | 0.531 | 0.597 | 0.543 | 0.757 | 0.617 | 0.544 | 0.667 | 0.617 | 0.589 | 0.666 |
| Adj. $R^2$ | 0.367 | 0.368 | 0.473 | 0.522 | 0.458 | 0.712 | 0.552 | 0.468 | 0.609 | 0.556 | 0.524 | 0.612 |
| RMSE | 1.535 | 0.501 | 0.402 | 1.594 | 0.527 | 0.343 | 1.264 | 0.454 | 0.339 | 1.264 | 0.431 | 0.339 |

A comparison of the three models reveals that the second form of the regression model is the best model for estimating Chl-*a* in deep and non-turbid water, with $R^2$ of 0.8 and RMSE of 0.3 compared to SARIMA,

NN-NAR, and NN-NARX with $R^2$/RMSE of 0.5/0.6, 0.4/0.3, and 0.3/5, respectively (Table 6). It is evident that Chl-*a* is dependent on the fluorescence of the algal species and SST which is one of the main regulators of algal growth in the sea. Therefore, this biological feature cannot depend only on its past values to forecast its future values and behavior. On the other hand, SARIMA has been found to be the best model for estimating FLH and SST in this location, with $R^2$/RMSE of 0.6/0.01 and 0.9/1.28, respectively. It can be claimed that FLH and SST can be modelled independently and the historical records of FLH and SST would be adequate to predict their future trends; they are not dependent on other parameters, unlike Chl-*a*, which is dependent on the physical properties of the sea. It is interesting to conclude that for deep water where a shallow bottom does not influence the satellite reading and where turbidity is low, a simple regression model can be used to model and forecast the Chl-*a* in this kind of ocean.

Table 6. Summary of the goodness of fit and RMSE of the NAR and NARX modelling.

| Parameter | NAR | | NARX | |
| --- | --- | --- | --- | --- |
| | $R^2$ | RMSE | $R^2$ | RMSE |
| Chl-*a* (3) | 0.774 | 2.859 | 0.522 | 2.022 |
| Chl-*a* (2) | 0.146 | 0.561 | 0.224 | 0.604 |
| Chl-*a* (1) | 0.367 | 0.529 | 0.270 | 0.546 |
| SST (3) | 0.205 | 0.007 | | |
| SST (2) | 0.204 | 0.007 | | |
| SST (1) | 0.528 | 0.006 | | |
| FLH (3) | 0.687 | 1.430 | | |
| FLH (2) | 0.713 | 2.287 | | |
| FLH (1) | 0.798 | 2.992 | | |

- Shallow (50-100m) and turbid water

Similarly, the time series modelling techniques have been validated for estimating the water quality parameters (Chl-*a*, FLH, and SST) in shallow and turbid water. It is interesting that regression is also found to be the best model for estimating the Chl-*a* in this kind of water, with $R^2$ and RMSE values of 0.6 and 0.4, respectively. However, it is the third form of regression described in Eq. 5 that is not only dependent on the co-current SST and FLH values but is also reliant on the past values of FLH (t-1) and SST (t-1). It is evident that as the water becomes shallower and more turbid, the complexity of the model increases.

Similar to the previous water type (deep and not-turbid), the SARIMA and neural network models are less effective at estimating the Chl-*a* in this water type (shallow and turbid), with $R^2$/RMSE values of 0.5/0.6 and 0.1/0.6, respectively. However, an improvement of the SARIMA model is observed in modelling Chl-*a* in this water type compared to the previous case of deep not-turbid water. Similarly, the SARIMA model is found to perform well in estimating SST and FLH in this type of water, with $R^2$/RMSE values of 0.98/1.05 and 0.7/0.006, respectively.

- Very shallow (<50m) and turbid water

In very shallow and turbid water, the neural network has shown the best performance in forescating Chl-*a* with $R^2$ and RMSE of 0.8 and 2.84. This is explained by the complexity of the model when it is applied in this complex situation of very shallow and turbid water. The third form of regression also performs well, with $R^2$ of 0.6 and RMSE of 1.26. However, SARIMA has shown very low performance, with $R^2$ of 0.3 and RMSE of 1.335.

As regards SST and FLH estimation, SARIMA has shown a great performance in estimating and predicting these two parameters in very shallow and turbid water. This proves that the SARIMA model can confidently be used to estimate the FLH and SST values of any water type varying from shallow to deep and turbid to non-turbid water.

- Overall validation

The different time series methods explored earlier have shown different sensitivities to the ocean parameters. For example, while the SARIMA method has shown the best prediction results for FLH and SST, the neural network method has shown the worst prediction results for SST. After comparing the three approaches (SARIMA, OLS regression, and neural network), one can claim that SARIMA is the best model for estimating FLH and SST in shallow-deep water and turbid-not turbid water. However, it is found that

Chl-*a* is perfectly predicted by the regression and neural network models as the water gets more complex in terms of turbidity and depth.

This finding might not work well in some cases due to the interferences of different water factors and the climatological features that creates unique features in each region. we also found it is difficult to find a model that can consider the peaks of Chl-*a*, which are not outliers because they are part of the data set. However, the SST data are much simpler to model than Chl-*a* data because of the clear seasonality and expected variability in this parameter. Likewise, the FLH data do not include the peaks, which makes the data more stationary. Table 7 shows the best model found for each dataset.

Table 7. The best model found for each parameter.

| Parameter | Best model | $R^2$ | RMSE |
|---|---|---|---|
| Chl-*a* (3) | Neural Network (NAR) | 0.7738 | 2.859 |
| Chl-*a* (2) | Regression (4) | 0.589 | 0.431 |
| Chl-*a* (1) | Regression (2) | 0.757 | 0.343 |
| FLH (3) | SARIMA(3,1,4) (1,1,1)$_{12}$ | 0.408 | 0.007 |
| FLH (2) | SARIMA(4,1,4) (1,1,1)$_{12}$ | 0.718 | 0.006 |
| FLH (1) | SARIMA(3,1,4) (1,1,1)$_{12}$ | 0.621 | 0.01 |
| SST (3) | SARIMA(2,1,2) (1,1,1)$_{12}$ | 0.978 | 0.564 |
| SST (2) | SARIMA(1,1,1) (1,1,1)$_{12}$ | 0.986 | 1.059 |
| SST (1) | SARIMA(4,1,4) (1,1,1)$_{12}$ | 0.983 | 1.289 |

5. Conclusions

Modeling the water quality parameters were done using three time series and machine learning approaches, which are univariate (SARIMA) and multivariate (regression) models and a neural network. First, the SARIMA model perfectly fitted the SST data in the three locations, which is reflected in the high coefficient of determination that was above 0.98. Similarly, high agreement was found between the real FLH and SARIMA forecast. However, the performance of SARIMA was poor in estimating and forecasting the Chl-*a* concentration in all water types (shallow-deep, and turbid-not turbid waters). Also, modeling Chl-*a* is more complex and the complexity of the model is dependent on water type. For the deep and less turbid waters, the regression model is able to forecast the Chl-*a*, but, as the water becomes shallower and its turbidity increases, the complexity of the model increases, and neural network is found to be the best for such complex cases. Overall, SARIMA is the best for modelling and forecasting FLH and SST parameters. However, neural network and/or regression models are the best models that can be used for modelling and forecasting chlorophyll *a* concentrations. The reason for this might be the complexity of the chlorophyll *a* dataset due to that is caused by the high level of spikes detected during the winter. Thus, we suggest examining more data points that are distributed at different areas to understand the spatial effect on the performance of the time series modelling.